\documentclass[sigconf, nonacm]{acmart}

\newcommand\vldbdoi{XX.XX/XXX.XX}
\newcommand\vldbpages{XXX-XXX}
\newcommand\vldbvolume{20}
\newcommand\vldbissue{X}
\newcommand\vldbyear{2027}
\newcommand\vldbauthors{\authors}
\newcommand\vldbtitle{\shorttitle}
\newcommand\vldbavailabilityurl{https://github.com/PanZaifeng/SmoothAgent}
\newcommand\vldbpagestyle{plain}

\usepackage{xurl}
\usepackage{amsmath}

\usepackage{graphicx}
\usepackage{algorithm}
\usepackage{algpseudocode}
\usepackage[dvipsnames]{xcolor}

\definecolor{codegreen}{rgb}{0,0.6,0}
\definecolor{codegray}{rgb}{0.5,0.5,0.5}
\definecolor{codepurple}{rgb}{0.58,0,0.82}
\definecolor{backcolour}{rgb}{0.95,0.95,0.92}
\definecolor{textblue}{rgb}{.2,.2,.7}
\definecolor{textred}{rgb}{0.54,0,0}
\definecolor{textgreen}{rgb}{0,0.43,0}
\definecolor{codered}{rgb}{201,72,12}

\usepackage[T1]{fontenc}
\usepackage[scaled=0.85]{beramono} 
\usepackage{listings}
\usepackage{multirow}
\usepackage{setspace}

\newcommand{\parahead}[1]{\textbf{#1}}
\def \SYS{SmoothAgent}

\begin{document}

\title{\SYS{}: Efficient Long-Horizon LLM-Based Agent Serving with Lookahead Context Engineering}

\author{Zaifeng Pan}
\affiliation{%
  \institution{University of California, San Diego}
}
\email{zapan@ucsd.edu}

\author{Qianxu Wang}
\affiliation{%
  \institution{University of California, San Diego}
}
\email{wqxwqx111@gmail.com}

\author{Zhengding Hu}
\affiliation{%
  \institution{University of California, San Diego}
}
\email{zhh068@ucsd.edu}

\author{Chang Chen}
\affiliation{%
  \institution{University of California, San Diego}
}
\email{chc278@ucsd.edu}

\author{Yue Guan}
\affiliation{%
  \institution{University of California, San Diego}
}
\email{y9guan@ucsd.edu}

\author{Yanbo Zhou}
\affiliation{%
  \institution{University of California, San Diego}
}
\email{yaz093@ucsd.edu}

\author{Steven Swanson}
\affiliation{%
  \institution{University of California, San Diego}
}
\email{sjswanson@ucsd.edu}

\author{Yufei Ding}
\affiliation{%
  \institution{University of California, San Diego}
}
\email{yufeiding@ucsd.edu}

\begin{abstract}

LLM-based agents execute multi-turn workflows with continuously growing contexts, where LLM calls are interleaved with tool invocations and environment feedback. To maintain model quality, modern agent frameworks rely on context engineering strategies such as offloading, reduction, and isolation to control the context length. However, these strategies introduce significant context transformation overhead: each transformation invalidates existing KV caches and triggers re-prefill, leading to increased time-to-first-token (TTFT).

In this paper, we identify that context transformations are segment-decomposable, where the transformation of a prefix is independent of future tokens. This property enables transformations to be executed ahead of time. Based on this insight, we propose a lookahead programming model that allows agent frameworks to express context transformations as asynchronous operations without modifying their execution logic. The runtime proactively executes these transformations and prepares transformed KV caches in advance, enabling direct context replacement without blocking.
We further design a lookahead-aware scheduler in LLM serving systems to support these asynchronous requests alongside latency-critical workloads with controlled interference. We implement our approach to support representative context engineering strategies and integrate it into existing agent frameworks and LLM serving systems. Experiments show that our approach effectively eliminates transformation overhead and reduces TTFT by up to 11.9$\times$.

\end{abstract}

\maketitle

\pagestyle{\vldbpagestyle}
\begingroup\small\noindent\raggedright\textbf{PVLDB Reference Format:}\\
\vldbauthors. \vldbtitle. PVLDB, \vldbvolume(\vldbissue): \vldbpages, \vldbyear.\\
\href{https://doi.org/\vldbdoi}{doi:\vldbdoi}
\endgroup
\begingroup
\renewcommand\thefootnote{}\footnote{\noindent
This work is licensed under the Creative Commons BY-NC-ND 4.0 International License. Visit \url{https://creativecommons.org/licenses/by-nc-nd/4.0/} to view a copy of this license. For any use beyond those covered by this license, obtain permission by emailing \href{mailto:info@vldb.org}{info@vldb.org}. Copyright is held by the owner/author(s). Publication rights licensed to the VLDB Endowment. \\
\raggedright Proceedings of the VLDB Endowment, Vol. \vldbvolume, No. \vldbissue\ %
ISSN 2150-8097. \\
\href{https://doi.org/\vldbdoi}{doi:\vldbdoi} \\
}\addtocounter{footnote}{-1}\endgroup

\ifdefempty{\vldbavailabilityurl}{}{
\vspace{.3cm}
\begingroup\small\noindent\raggedright\textbf{PVLDB Artifact Availability:}\\
The source code, data, and/or other artifacts have been made available at \url{\vldbavailabilityurl}.
\endgroup
}

\section{Introduction}
\label{sec:introduction}

\begin{figure}
    \centering
    \includegraphics[width=\linewidth]{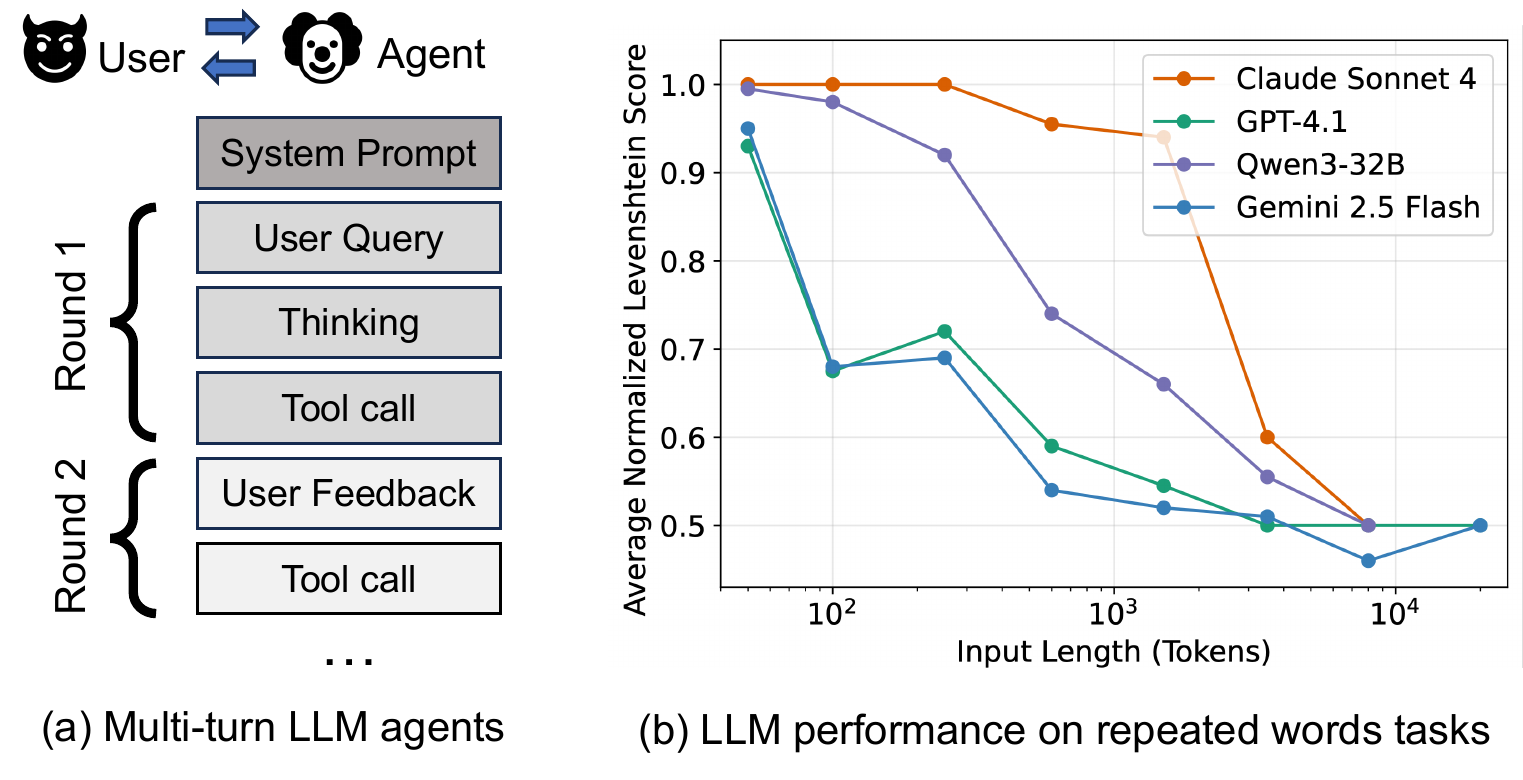}
    \caption{Without context engineering, the contexts of LLM agents will accumulate rapidly through rounds, leading to the context rot phenomenon. The experimental results in (b) are reported by Chroma~\cite{hong2025context}.}
    \Description{Context growth over agent rounds and the resulting context rot effect.}
    \label{fig:context_rot}
\end{figure}

Large language model (LLM)-based agents~\cite{yao2023react,shinn2023reflexion,lin2023swiftsage,hong2023metagpt,yang2024swe,anthropic_agent,liu2025supporting,luo2026data,giannakouris2025lambda,lu2025adda} are increasingly deployed in real-world applications, where they execute multi-turn workflows with interleaved model inference, tool usage, and user interaction, as shown in Figure~\ref{fig:context_rot}(a).
In such scenarios, the context continuously accumulates conversation history, tool outputs, and intermediate states, quickly becoming a central factor for both model quality and system efficiency. 
As Figure~\ref{fig:context_rot}(b) shows, prior work has identified \textit{context rot}~\cite{hong2025context}, where model reliability degrades as the context grows, even within the supported context window.
Meanwhile, longer contexts also incur higher computational cost and GPU memory usage, making effective context management essential for production LLM agents.

\begin{figure}
    \centering
    \includegraphics[width=\linewidth]{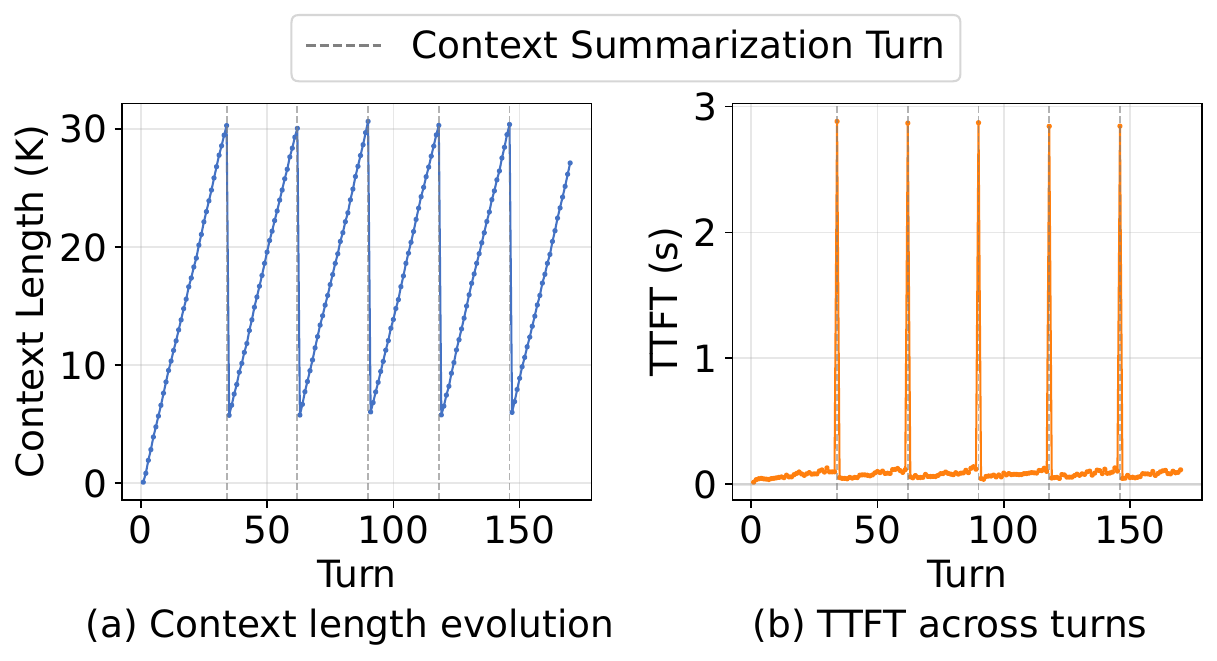}
    \caption{By periodically applying context summarization, LLM agents effectively control the context length, but introduce significant context transformation overhead.}
    \Description{Context summarization controls context length but introduces TTFT overhead at transformation points.}
    \label{fig:context_ttft}
\end{figure}

To mitigate context rot, production agents~\cite{openclaw,Liu_LlamaIndex_2022,autogen,langchain,ji2025manus} adopt a variety of context engineering strategies to control the context length.
Common context engineering techniques include reduction~\cite{langchain_context_engineering}, which truncates or summarizes historical information; offloading~\cite{ji2025manus}, which moves parts of the context such as tool outputs to file systems; and isolation~\cite{langchain_context_engineering,huggingface_open_deep_research}, which launches sub-agents with separate and clean contexts. 
As the number of interaction turns increases, the accumulated context grows continuously, while these strategies periodically reduce or reorganize the context to maintain a manageable size. As shown in Figure~\ref{fig:context_ttft}(a), triggering operations such as offloading or summarization effectively controls context growth over time.
Context engineering has become a fundamental component of agent harness development and is critical to the effectiveness of agent frameworks.

\begin{figure}
    \centering
    \includegraphics[width=\linewidth]{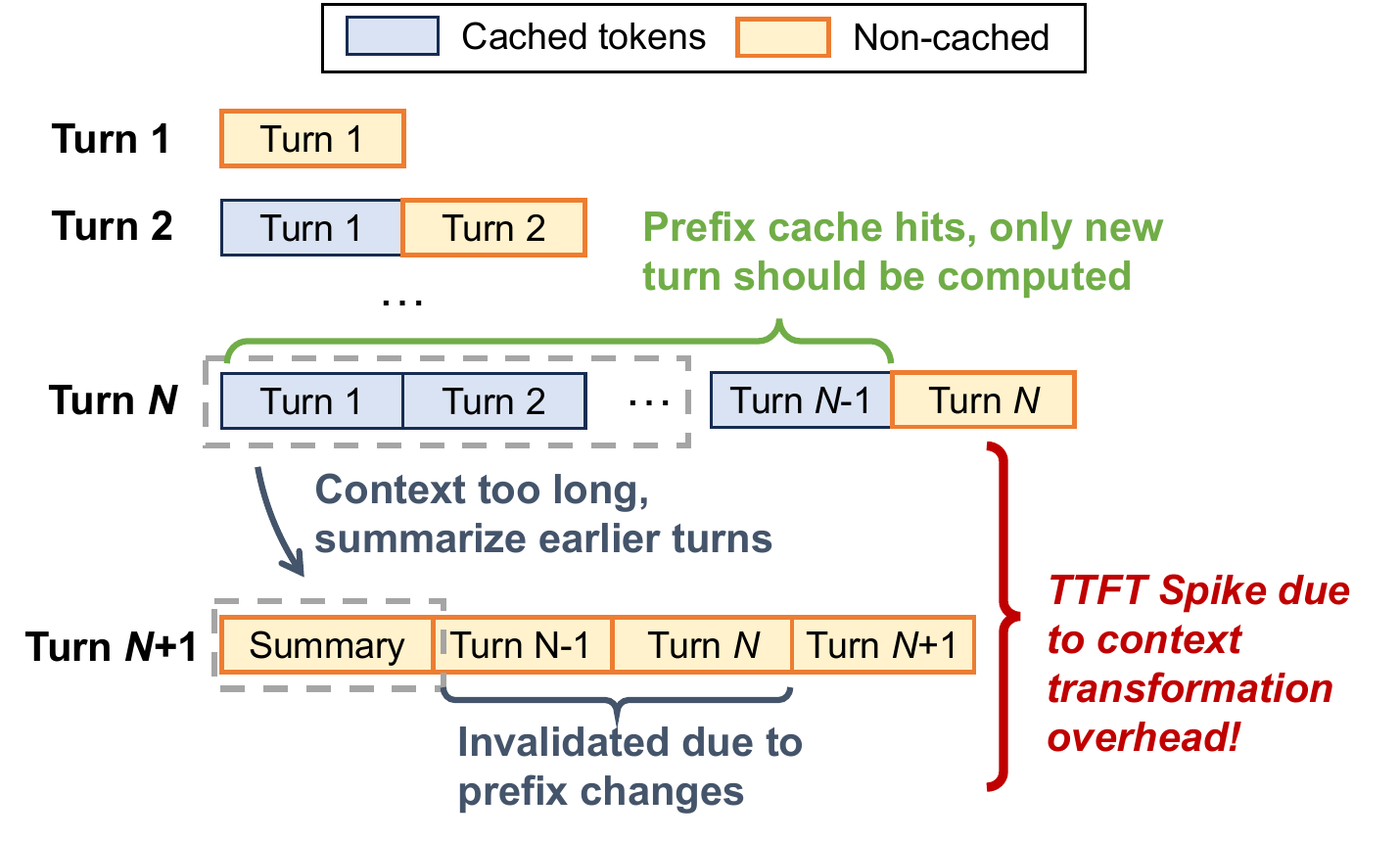}
    \caption{An illustrative example of context transformation overhead, including KV cache invalidation and additional transformation cost, which leads to TTFT spikes.}
    \Description{A multi-turn agent context reuses cached prefix tokens across turns. When earlier turns are summarized, the prefix changes, existing KV cache entries become invalid, and the next response incurs a TTFT spike.}
    \label{fig:intro_example}
\end{figure}

However, although context engineering improves agent accuracy and robustness, it introduces a new performance challenge. From a system perspective, these techniques fundamentally perform context transformation, where the original context is mapped to a new representation.
This transformation invalidates the existing KV cache, forcing the system to recompute it through a new prefill stage. As a result, each transformation incurs additional latency, which we refer to as \textit{context transformation overhead}.
Figure~\ref{fig:intro_example} illustrates this overhead with a summarization example that rewrites earlier turns and changes the context prefix.
As illustrated in Figure~\ref{fig:context_ttft}(b), when such transformations are triggered, the time-to-first-token (TTFT) increases significantly, particularly affecting tail latency.
In latency-sensitive applications such as real-time simulation~\cite{ren2025simworld,zhuang2025simworld}, this increase in tail TTFT can lead to unacceptable violations of service-level objectives (SLOs), severely degrading user experience.

To address the context transformation overhead, we first analyze commonly used context engineering techniques and make a key observation: most context transformations exhibit a \textit{segment-decomposable} property.
Intuitively, this property implies that the transformation of a context can be decomposed into transformations over its individual segments, such that applying the transformation to each segment independently and then concatenating the results yields the same outcome as transforming the entire context at once.
This means that the transformation of each segment is independent of the tokens that follow it, enabling opportunities to perform transformations in advance.
We formalize this segment-decomposable property in Section~\ref{sec:lookahead} and show that it holds for various context engineering strategies.

Based on this insight, we propose a \textit{lookahead programming model} that eliminates context transformation overhead by decoupling transformation from the critical path. The key idea is to maintain a lookahead context alongside the working context and asynchronously compute the KV cache of transformed contexts in advance. When a context transformation is required, the system can directly replace the KV cache without performing a new prefill.
For example, in some agents that offload all tool outputs to file systems once a context limit is reached, the transformed context after each tool call is independent of future tokens. This allows the system to precompute the corresponding KV cache in the lookahead stream after each tool call and seamlessly switch to it when reaching context limit.
We demonstrate how various context engineering techniques can be expressed within this programming model and benefit from reduced overhead in Section~\ref{sec:lookahead}.

While lookahead computation reduces transformation latency, it introduces potential interference with foreground requests.
Lookahead requests may compete with the main context generation or with requests from other users in a shared serving environment.
To address this challenge, we design a \textit{lookahead-aware request scheduler} that minimize interference under both prefill-decode aggregation and disaggregation settings. Specifically, we split lookahead requests into dynamic chunks and propose SLO-aware lookahead and latency-critical request co-batching strategies. These ensure that lookahead computation is performed only when it does not compromise the latency guarantees of other requests.
In addition, we introduce a promotion interface in the programming model that prioritizes lookahead requests when a context transformation is imminent, ensuring timely availability of transformed KV caches.

We implement our approach by integrating optimized lookahead-based context engineering strategies into popular agent frameworks, including MiniAgent~\cite{miniagent}, OpenClaw~\cite{openclaw}, LangChain~\cite{langchain}, LlamaIndex~\cite{Liu_LlamaIndex_2022}, and AutoGen~\cite{autogen}. Our evaluation demonstrates that the proposed methods significantly improve tail TTFT by effectively hiding context transformation overhead, leading to a more smooth TTFT profile.

In summary, this paper makes the following contributions:

\begin{itemize}
\item We identify and characterize the context transformation overhead introduced by context engineering in LLM agent workloads.
\item We formalize the segment-decomposability property of commonly used context transformations.
\item We propose a lookahead programming model that generalizes across context engineering strategies and eliminates transformation overhead, along with SLO-aware scheduling techniques to mitigate request interference.
\item We conduct comprehensive experiments across strategies and agent frameworks, demonstrating up to 11.9$\times$ improvement in transform-point TTFT.
\end{itemize}

\section{Background and Motivation}
\label{sec:background}

\parahead{LLM agents and context engineering.}
LLM-based agents~\cite{yao2023react,yang2024swe,anthropic_agent,zeng2026glm} have emerged as a dominant paradigm for solving complex, multi-step tasks, and are widely adopted in both research and production systems~\cite{wang2025openhands,hong2023metagpt,huggingface_open_deep_research,yao2022webshop,qin2025ui,anthropic2025research,claudecode}. Instead of a single forward pass, agents iteratively interact with the environment through interleaved LLM calls, tool invocations, and human feedback. Each step produces new observations that are appended to the context, forming a growing execution trace over time.

As a result, agent workloads naturally consist of long, multi-turn sessions with rapidly expanding context. This growth introduces two key challenges. First, the context may exceed the model's context window, forcing the system to discard, compress, or externalize information. Second, even within the limit, models often suffer from \textit{context rot}~\cite{hong2025context}, where performance degrades as the context becomes longer and noisier, as illustrated in Figure~\ref{fig:context_rot}(b).

These challenges make context management a first-class concern in agent systems. Recent studies and production deployments show that agent performance depends critically on how the context is constructed, maintained, and transformed. This has led to the emergence of \textit{context engineering}~\cite{ji2025manus,anthropic_context_engineering,langchain_context_engineering} as a fundamental component of modern agent frameworks.

To manage context growth, agent frameworks apply a range of context engineering strategies~\cite{ji2025manus,anthropic_context_engineering,langchain_context_engineering,liu2025deepseek,context_for_deepagents,zeng2026glm}. For example, some systems offload past interactions or tool outputs to external storage, others summarize long histories to reduce context length, and many isolate sub-tasks into separate sub-agent contexts with dedicated prompts. These transformations help control context size and improve robustness, but they also introduce execution overhead. Each transformation modifies the input prefix and requires recomputing the KV cache, which lies on the critical path of inference and can lead to noticeable TTFT spikes, as shown in Figure~\ref{fig:context_ttft}.

\parahead{Agent serving systems.}
The serving stack for LLM agents typically consists of two loosely coupled layers. At the frontend, agent frameworks~\cite{claudecode,openclaw,langchain,Liu_LlamaIndex_2022,miniagent} define the agent harness, including control flow, tool usage, and context management policies. These frameworks focus on expressiveness and task-solving capability, enabling developers to compose complex agent behaviors.

At the backend, LLM serving systems such as vLLM~\cite{vllm} and SGLang~\cite{sglang} optimize model execution efficiency. They employ techniques such as batching and scheduling~\cite{yu2022orca,agrawal2024taming,zhong2024distserve,patel2024splitwise,zheng2024batchllm}, kernel optimization~\cite{ye2025flashinfer,dao2022flashattention,kamath2025pod,dong2024flex,pan2025fasttree}, and KV cache management~\cite{vllm,qin2025mooncake,abhyankar2024infercept,yao2025cacheblend} to improve throughput and reduce latency across concurrent requests.

In this work, we improve the execution efficiency of agent systems under context engineering. At the agent framework level, we provide a lookahead programming model that enables context transformations to be expressed in a form that can be executed asynchronously without changing their logic. At the LLM serving system level, we optimize the scheduler to support these lookahead requests, allowing them to run alongside regular requests with controlled interference.
\section{Lookahead Programming Model for Context Transformation}
\label{sec:lookahead}

\subsection{Segment-Decomposable Transformation}

In this section, we formalize a common structural property that enables more efficient context engineering execution.
In particular, we view context engineering as a transformation process over the accumulated context. Let $C$ denote the current context of the agent, and let $T(\cdot)$ denote a transformation function that is triggered when certain conditions are met, such as the context length exceeding a predefined threshold. The transformed context $T(C)$ is then used for subsequent LLM invocations to maintain generation quality.

\begin{figure}
    \centering
    \includegraphics[width=0.99\linewidth]{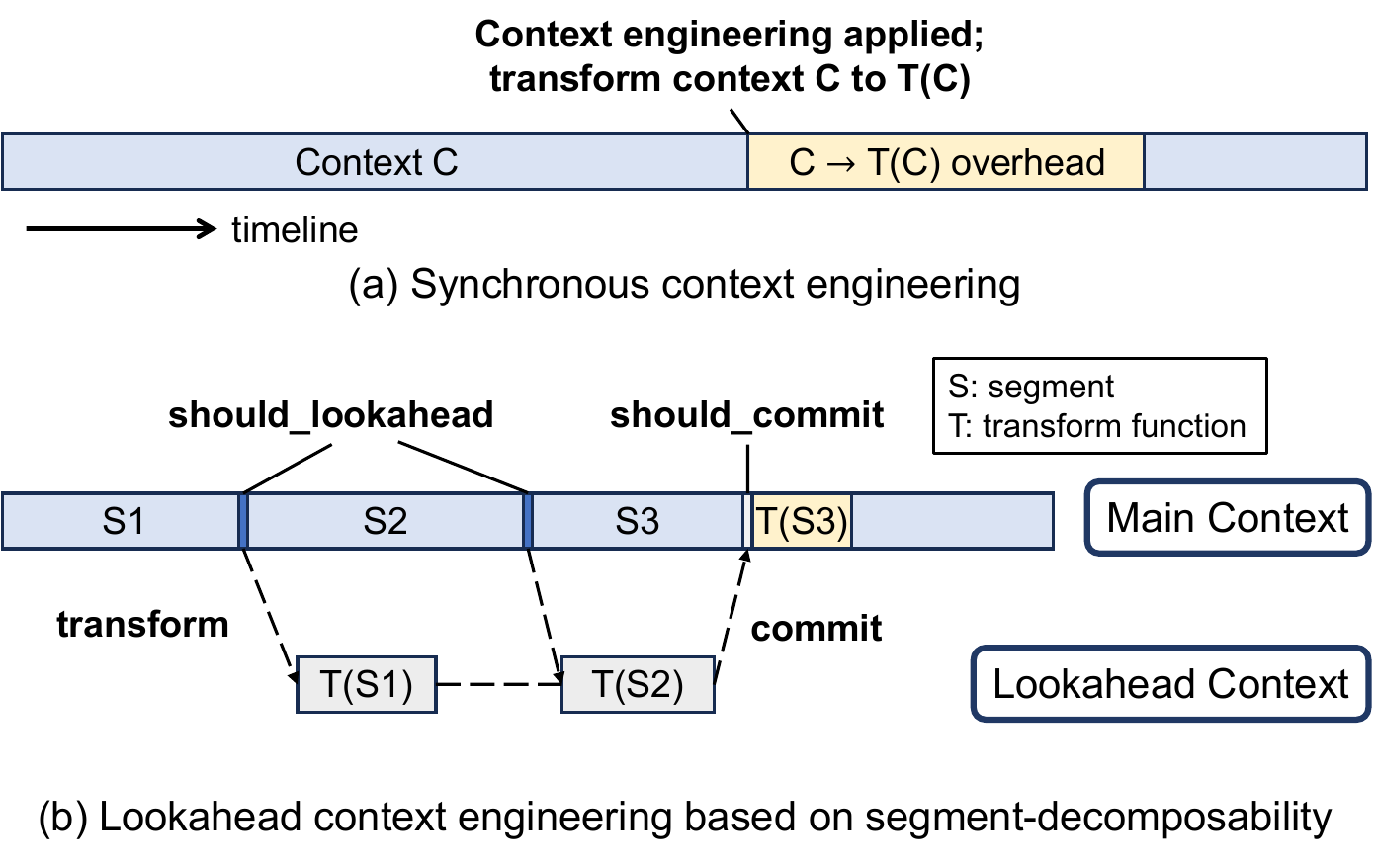}
    \caption{Execution timeline of the lookahead programming model. The main stream proceeds normally while the lookahead stream asynchronously computes the transformed segments' KV cache in advance. At commit time, the pre-computed result is spliced.}
    \Description{Timeline showing normal agent execution alongside asynchronous lookahead context transformation and commit.}
    \label{fig:lookahead_model}
\end{figure}

In existing systems, such transformations are typically executed synchronously on the critical path of generation, introducing non-trivial overhead. First, applying $T$ invalidates the KV cache of the original context, requiring recomputation for the transformed input. Second, the transformation itself incurs additional cost, which may involve file system operations, external storage access, or auxiliary model calls such as summarization. As a result, synchronous transformation increases latency, particularly affecting TTFT in production systems.
This motivates the need to execute context transformations ahead of time, without blocking ongoing generation.

To understand when such asynchronous execution is possible, consider the example of context offloading. As illustrated in Figure~\ref{fig:offloading}, when the context exceeds a threshold, agent frameworks may offload certain components, such as tool outputs, to file systems, thereby reducing the working context length. In this process, the transformation applied to each tool output is independent of others. Once a tool output is produced, its transformed representation is fully determined and does not depend on future tokens or subsequent interactions. Therefore, its transformation can be computed immediately, rather than waiting for the global offloading trigger.

This observation motivates the notion of \emph{segment-decomposability}. Let the context $C$ be partitioned into a sequence of segments $\{ S_1,\allowbreak S_2,\allowbreak \dots,\allowbreak S_n \}$. A transformation function $T$ is segment-decomposable if there exists such a partition where each segment can be transformed independently, and the overall transformation can be expressed as
\[
T(C) = T(S_1) \;\Vert\; T(S_2) \;\Vert\; \cdots \;\Vert\; T(S_n),
\]
where $\Vert$ denotes concatenation. Under this property, each segment can be transformed as soon as it becomes available, enabling asynchronous execution in advance and avoiding increases in TTFT.

In Sections~\ref{subsec:strategy_start}-\ref{subsec:strategy_end}, we show that widely used context engineering strategies, including offloading, reduction, and isolation, naturally exhibit this property, which enables their transformation overhead to be hidden via lookahead execution.

\subsection{Lookahead Programming Model}
\label{subsec:programming_model}

Building on segment-decomposability, we design a \textit{lookahead programming model} that expresses context engineering strategies in a form that exposes transformation opportunities to the runtime. 
The key idea is to maintain a \emph{lookahead context} alongside the main working context, as illustrated in Figure~\ref{fig:lookahead_model}. 
%
The model consists of a strategy interface together with a runtime that asynchronously executes lookahead transforms and finally commits their results into the main context. We present their design in the rest of this section.

\parahead{State abstractions.}
The model maintains two state objects corresponding to the working context and the lookahead context, as shown in Listing~\ref{list:lookahead}.
\texttt{MainState} captures the current working context, including the full message list and per-strategy metadata such as token counts and limits, which are used to determine when to trigger different actions.
\texttt{LookaheadState} tracks the progress of lookahead execution. It contains (1) \texttt{transformed}, the accumulated transformed segments constructed incrementally across processed segments; (2) \texttt{last\_segment\_end}, which marks the end of the prefix that has already been incorporated into the lookahead state; and (3) \texttt{strategy\_data}, a per-strategy mutable state used to store auxiliary information such as file handles for offloading or intermediate summaries.

\parahead{Strategy interface.}
As shown in Listing~\ref{list:lookahead}, each context engineering strategy implements the following methods:

\begin{itemize}
\item \texttt{should\_lookahead() $\to$ bool}:
invoked at segment boundaries to determine whether the newly available segment should be transformed asynchronously. Different strategies instantiate this condition differently.

\item \texttt{transform()}:
incrementally updates the transformed prefix using the current segment. The function mutates \texttt{la\_state} to reflect updated progress. By segment-decomposability, the transformation depends only on the current segment and the existing lookahead state, and is independent of any future tokens.

\item \texttt{should\_commit() $\to$ bool}:
determines whether the precomputed transformed prefix should be committed into the working context. This condition is typically triggered when the context exceeds the token limit.

\item \texttt{should\_promote() $\to$ bool}:
provides an early signal to increase the scheduling priority of pending lookahead requests. By default, this condition aligns with \texttt{should\_commit}, but strategies may override it to promote lookahead execution before the commit point, reducing the likelihood of misses.
\end{itemize}

\begin{minipage}[H]{0.95\linewidth}
\centering
\begin{lstlisting}[caption=Strategy interface of the programming model,label=list:lookahead]
@dataclass
class MainState:
    messages: list[Message]
    strategy_data: Dict

@dataclass
class LookaheadState:
    transformed: List[Message]
    last_segment_end: int
    strategy_data: Dict

class LookaheadStrategy(ABC):
    def __init__(self):
        self.main = MainState(...)
        self.la = LookaheadState(...)
    
    @abstractmethod
    def should_lookahead(self) -> bool: ...
    
    @abstractmethod
    def transform(self): ...
    
    @abstractmethod
    def should_commit(self) -> bool: ...
    
    def should_promote(self) -> bool:
        return self.should_commit()
\end{lstlisting}
\end{minipage}

\parahead{Execution runtime.}
The runtime maintains two concurrent streams, as illustrated in Figure~\ref{fig:lookahead_model}. The \textit{main stream} executes the agent loop as usual, while the \textit{lookahead stream} processes segments asynchronously.

At each segment boundary, the runtime invokes the function \texttt{should\_lookahead}. If triggered, it extracts the newly available segment and submits a \texttt{transform} call as a lookahead request. Although \texttt{transform} is defined over a single segment, its execution introduces two challenges. First, LLM calls within \texttt{transform} require KV cache construction that depends on the previously transformed prefix. Second, the lookahead state must be updated incrementally to reflect the progressive construction of the transformed context. To preserve this dependency and for simplicity, lookahead requests are enqueued and executed in order, so that each request observes the appropriate prefix state. All LLM requests issued within the \texttt{transform} function are marked as \emph{lookahead requests} and scheduled outside the latency-critical path. We discuss their scheduling in Section~\ref{sec:colocate}.

When \texttt{should\_commit} fires, the runtime attempts to commit the lookahead result by splicing the transformed prefix into the working context. If the corresponding lookahead computation has completed, the runtime directly replaces the prefix in the working context with the transformed result, thereby avoiding blocking transformation on the critical path. Otherwise, the runtime promotes the priority of pending lookahead requests so that the scheduler can expedite their execution. Such promotion can also happen proactively before the commit point if \texttt{should\_promote} is satisfied.

Listing~\ref{list:agent_loop} shows how the programming model integrates into a standard agent loop. The runtime encapsulates the complexity of lookahead scheduling and result management, while developers only need to implement the strategy interface and insert lightweight hooks into the agent loop.

\begin{minipage}[H]{0.95\linewidth}
\centering
\begin{lstlisting}[caption=Agent loop with lookahead context engineering,label=list:agent_loop]
strategy = MyLookaheadStrategy(...)
runtime  = LookaheadRuntime(strategy)

messages = [system_prompt, user_query]
while not task_done:
    # Commit hook: if context budget is exceeded, await
    # lookahead transforms and promote if necessary.
    # Then splice the transformed prefix:
    #   main.messages <= la.transformed
    #     ++ T(messages[la.last_segment_end:])
    messages = runtime.on_commit(messages)

    response = llm.generate(messages)
    messages.append(response)
    if response.has_tool_call:
        obs = execute_tool(response.tool_call)
        messages.append(obs)

    # Segment hook: extract new segment since last
    # trigger and enqueue a background transform.
    # Completed result is appended to la.transformed.
    runtime.on_segment_boundary(messages)
\end{lstlisting}
\end{minipage}

In the following sections, we instantiate this programming model with commonly used context engineering strategies, showing how each strategy satisfies segment-decomposability and maps naturally onto the lookahead execution model.

\subsection{Offloading}
\label{subsec:strategy_start}

\begin{figure}
    \centering
    \includegraphics[width=0.99\linewidth]{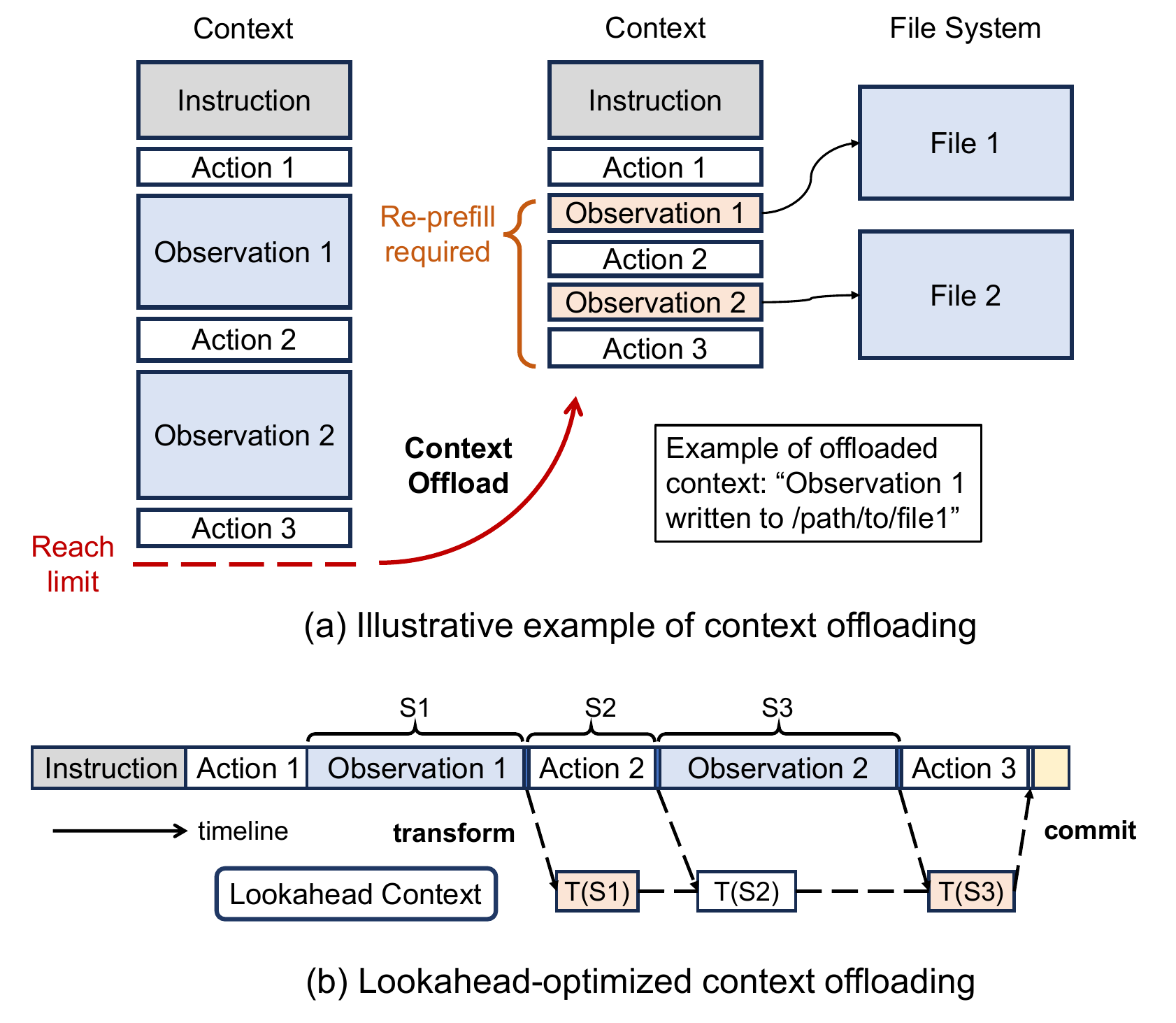}
  \caption{Offloading transforms bulky observations (e.g., tool outputs) into compact references to external storage. Since each completed observation can be rewritten independently, the transformed KV cache can be prepared ahead of time.}
    \Description{Context offloading replaces bulky observations with compact external-storage references.}
    \label{fig:offloading}
\end{figure}

Context offloading~\cite{ji2025manus,context_for_deepagents} rewrites the context by moving observations (e.g., tool execution results, environment states) to external storage and replacing them with lightweight references. As illustrated in Figure~\ref{fig:offloading}(a), when the context approaches a predefined limit, earlier observations are written to the file system. Their in-context representations are replaced with pointers such as ``Observation 1 written to /path/to/file1'', which triggers a new blocking prefill request. This design preserves the full information in a restorable form while reducing the active context length. Unlike lossy techniques such as summarization, offloading maintains exact fidelity and allows the agent to read file contents when needed.

Offloading fits naturally into our segment-decomposable abstraction. The decision to offload depends only on the current context up to a segment boundary (e.g., after an observation), and each segment can be transformed independently of future segments. This property enables the system to incrementally construct transformed prefixes ahead of time and overlap transformation with ongoing execution.

As shown in Figure~\ref{fig:offloading}(b), \texttt{should\_trigger} is evaluated after each segment is appended to determine whether the current segment should be offloaded. Once triggered, \texttt{transform} incrementally rewrites the prefix by offloading completed segments to external storage and replacing them with references in the lookahead stream, while updating the lookahead state to track the materialized outputs. In parallel, \texttt{should\_commit} is evaluated as the main execution progresses; when the context reaches the offloading point, the system replaces the corresponding prefix in the main context with the transformed prefix produced by the lookahead execution.

By preparing the transformed prefix in advance, the system avoids performing the transformation and re-prefill synchronously on the critical path, effectively hiding the overhead of context offloading.

\subsection{Reduction}

Context reduction~\cite{langchain_context_engineering,context_for_deepagents,li2023compressing} controls context growth by discarding or compressing less relevant history while retaining information that is most useful for future reasoning. Two representative forms are truncation and summarization.

\begin{figure}
    \centering
    \includegraphics[width=0.92\linewidth]{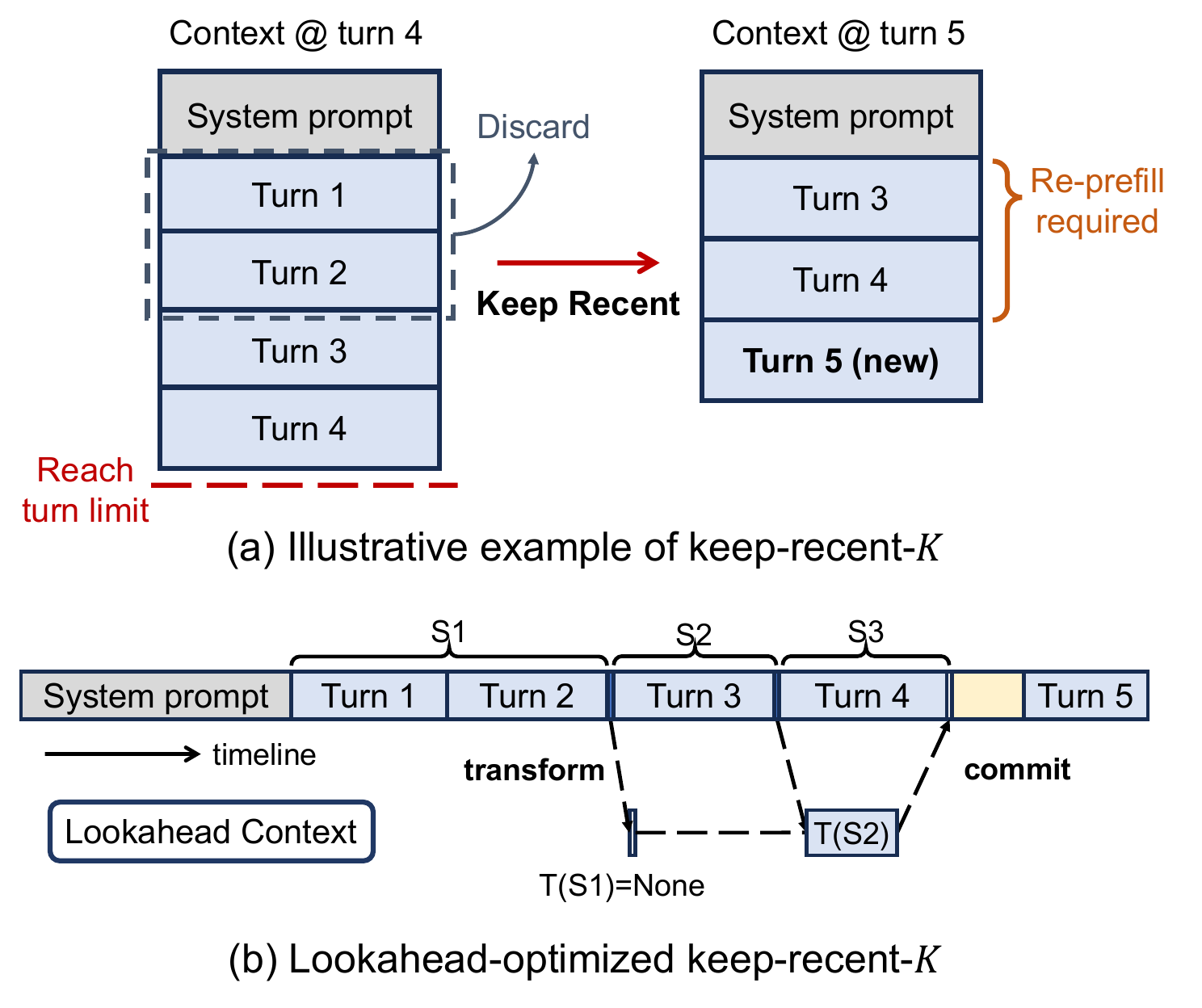}
  \caption{Keep-recent-$K$ reduction keeps only the most recent $K$ context segments. The transform is local to the retention boundary, which meets the segment-decomposable property.}
    \Description{Keep-recent-K reduction drops old context segments while retaining the newest segments.}
    \label{fig:sliding}
\end{figure}

\parahead{Truncation.}
Truncation strategies directly discard historical contents to control context growth. The most common and simple strategy is keep-recent-$K$~\cite{liu2025deepseek,zeng2026glm}, which retains only the most recent $K$ turns while discarding older ones. As shown in Figure~\ref{fig:sliding}(a), when a new turn is appended and the number of turns exceeds the limit, the oldest turns are removed. Although the contents in Turns 3 and 4 do not change, a new prefill is still required to rebuild the KV cache for subsequent generations. This approach is simple yet effective, as recent interactions are typically the most relevant for next-step reasoning in agent workloads~\cite{liu2024lost}.

The keep-recent-$K$ example also fits into the segment-decomposable abstraction. Each turn can be treated as a segment, and whether a segment is retained depends only on its relative position within the window. Since removal decisions do not depend on future segments, the transformed prefix can be constructed incrementally.

As illustrated in Figure~\ref{fig:sliding}(b), \texttt{should\_trigger} is evaluated after each new turn to determine whether the window size exceeds the limit. Once triggered, \texttt{transform} incrementally drops outdated segments and constructs the truncated prefix, computing the corresponding KV cache in advance. When the window boundary shifts, the system detects the commit signal and replaces the corresponding prefix with the truncated version produced by lookahead execution.

\begin{figure}
    \centering
    \includegraphics[width=0.99\linewidth]{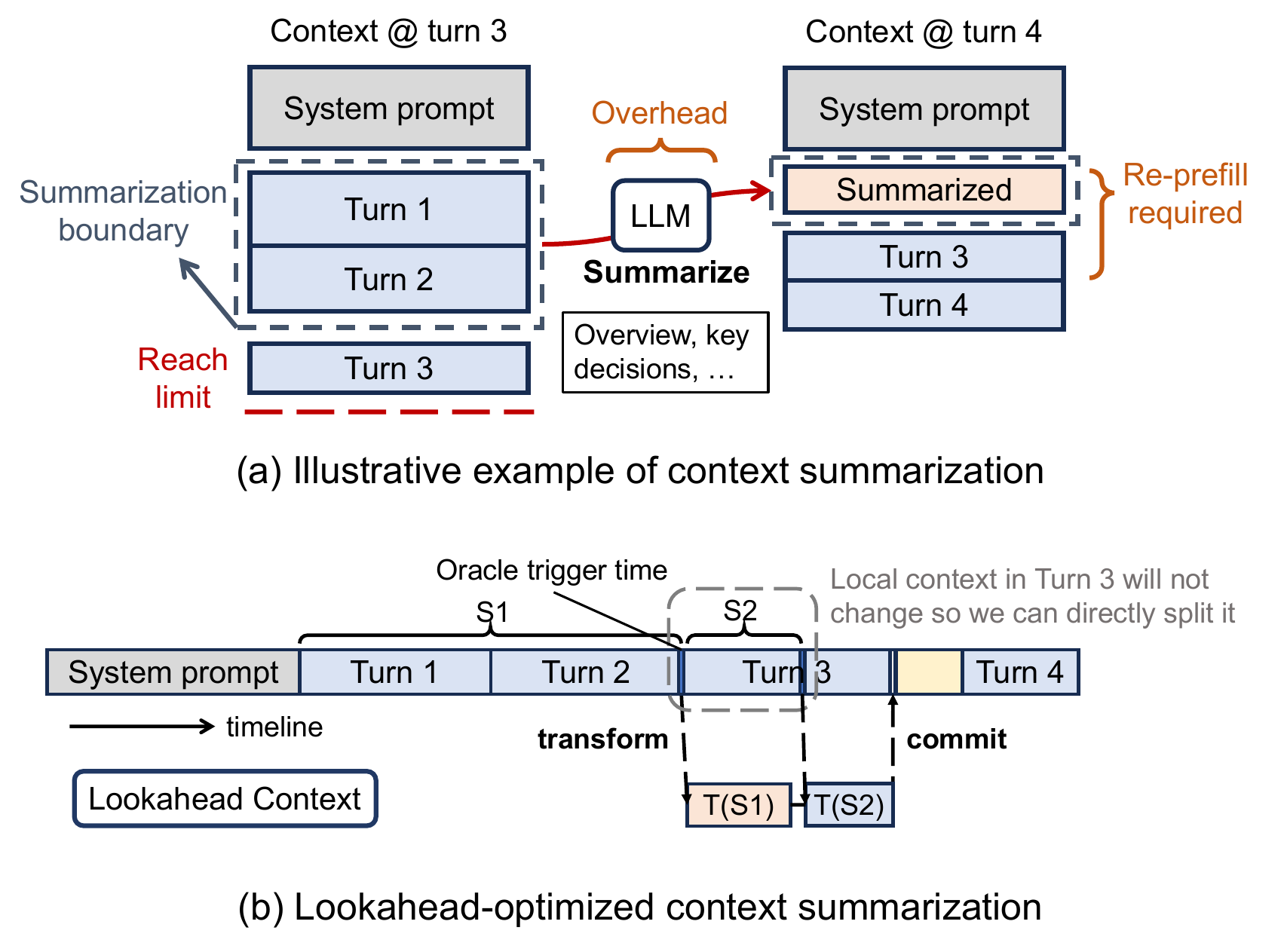}
  \caption{Summarization compresses the prefix into a compact synopsis. The summary depends only on the completed history, which makes it amenable to lookahead execution.}
    \Description{Summarization compresses completed context history into a shorter synopsis.}
    \label{fig:summarization}
\end{figure}

\parahead{Summarization.}  
Another form of reduction replaces a span of past context with a compact summary generated by an LLM. As shown in Figure~\ref{fig:summarization}(a), when the context reaches a predefined limit, a subset of earlier turns is summarized into a shorter representation (e.g., key decisions or high-level descriptions), which is then inserted back into the context. This transformation reduces context length while preserving approximate semantic information.

Summarization also exhibits segment-decomposability, but with a coarser granularity. In practice, agents typically preserve recent turns verbatim and only summarize earlier history to maintain fidelity. As a result, summarization only applies to the completed segments before the summarization boundary, making it independent of future context.

Determining the exact summarization boundary may require observing the full context (e.g., when summarizing all but the most recent $N$ turns). To enable lookahead execution, we use a \textit{soft trigger threshold} that is slightly smaller than the actual summarization limit. When this soft threshold is reached, \texttt{should\_trigger} initiates a lookahead summarization request based on the current prefix. Although the resulting boundary may not exactly match the final one, the discrepancy is bounded by the gap between the soft and hard thresholds. In practice, this introduces negligible impact when the total context is long, and may even improve fidelity by summarizing slightly less content.

\subsection{Isolation}
\label{subsec:strategy_end}

\begin{figure}
    \centering
    \includegraphics[width=0.92\linewidth]{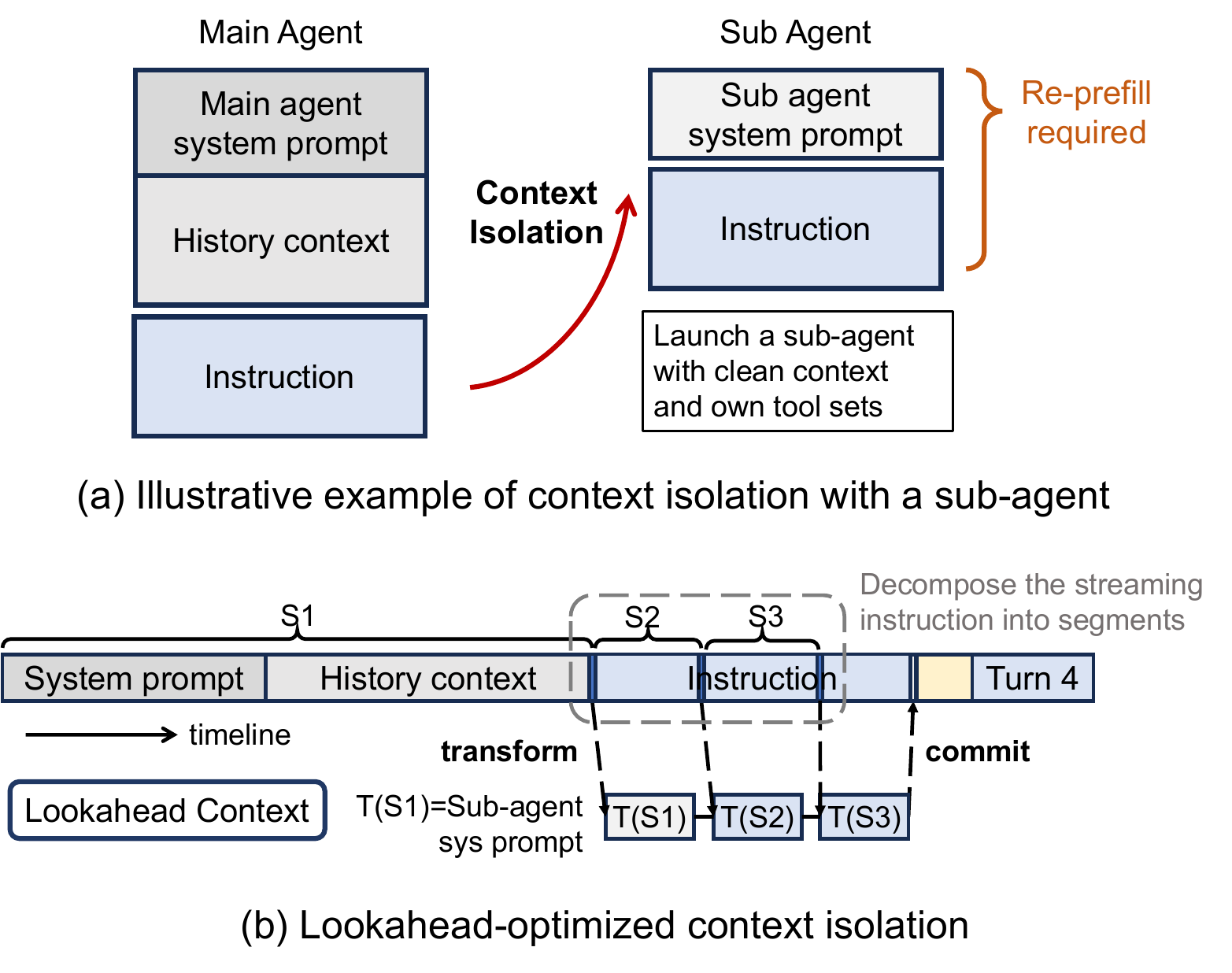}
  \caption{Context isolation launches a sub-agent with a clean context. By decomposing instruction generation into segments, the sub-agent context can be prepared ahead of time.}
    \Description{Sub-agent isolation creates a separate clean context and incrementally prepares it during instruction generation.}
    \label{fig:isolation}
\end{figure}

Context isolation~\cite{anthropic_context_engineering,langchain_context_engineering} separates a task into a new execution context, typically by launching a sub-agent with its own system prompt and tool set. This design helps reduce interference from irrelevant history, improves modularity, and allows the sub-agent to operate with a specialized prompt and tool configuration tailored to the delegated task.
As shown in Figure~\ref{fig:isolation}(a), when the main agent delegates a task, it first generates an instruction describing the sub-task, and then switches to a clean context for the sub-agent. This new context consists of the sub-agent's system prompt and the generated instruction, and requires a fresh prefill before execution.
This process can be viewed as a context transformation that replaces the main-agent prefix with a new sub-agent context.

Isolation can also be expressed within the segment-decomposable abstraction. We treat the point where the main agent decides to delegate a sub-agent as the first segment boundary, and transform it into the sub-agent system prompt in advance.
After that, the main agent generates instructions for the sub-agent. This instruction generation, however, is not naturally decomposed for context transformation. Treating it as a monolithic output delays sub-agent context construction until decoding completes, leading to performance for long instructions.

To address this issue, we reformulate instruction generation as a \textit{streaming} process and partition it into fine-grained segments. Each segment corresponds to a chunk of newly generated tokens and depends only on the previously generated prefix, making it compatible with segment-decomposability. This enables the system to incrementally construct the sub-agent context alongside decoding, rather than waiting for the full instruction.
Since instruction generation is a relatively long-running auto-regressive decoding process, it creates sufficient opportunity to prepare the sub-agent KV cache incrementally. As a result, most of the prefill cost can be amortized and hidden before the context switch occurs.
FlashAgents~\cite{fang2026flashagents} also leverages similar streaming ideas to accelerate multi-agent workflows.

As illustrated in Figure~\ref{fig:isolation}(b), \texttt{should\_trigger} is activated when the system decides to launch a sub-agent. During instruction generation, the decoding stream is partitioned into segments (S2, S3), each corresponding to a chunk of the instruction. For each segment, \texttt{transform} incrementally constructs the target sub-agent context by appending the generated instruction chunk to the sub-agent system prompt and preparing the corresponding KV cache. Since instruction generation is typically slower than prefilling, most of the sub-agent context can be prepared in advance and hidden from the decoding process.

Finally, \texttt{should\_commit} is evaluated when the instruction generation completes. At this point, the system replaces the main-agent context with the constructed sub-agent context and performs a final short prefill to complete the transition. By preparing the majority of the sub-agent context in advance, the system significantly reduces the blocking prefill overhead of context isolation.

\section{Lookahead-Aware Request Scheduler}
\label{sec:colocate}

Co-serving lookahead requests with other requests introduces interference that can degrade serving latency. On prefill-only instances~\cite{zhong2024distserve,patel2024splitwise,qin2025mooncake}, lookahead requests consume service time and increase the queuing delay of other requests. In prefill-decoding (PD) co-located deployments~\cite{agrawal2024taming,holmes2024deepspeed,kamath2025pod}, including lookahead chunks in a batch increases its execution time, delaying subsequent iterations. As a result, both time-to-first-token (TTFT) and time-between-tokens (TBT) can be negatively affected.

These effects are particularly problematic for requests on the critical path of agent execution, including prefills and decode iterations in the main context. Their latency directly determines TTFT and TBT, and we refer to them as \textit{latency-critical} (LC) requests.

To mitigate such interference, \SYS{} treats lookahead requests as \textit{best-effort} (BE) jobs. Lookahead requests have inherently looser timing requirements: they only need to complete before their corresponding commit point to fully realize their benefit, and can otherwise fall back to synchronous execution without incurring additional overhead compared to the baseline. Therefore, BE jobs are executed only when sufficient slack exists and must never degrade the latency guarantees of LC requests. Concretely, LC requests always take priority, and lookahead computation is admitted only if it does not violate TTFT or TBT constraints.

Enforcing this policy requires accurately estimating the latency impact of co-batching lookahead and LC requests. We first present a performance model for batch execution latency, and then describe SLO-aware co-batching strategies for both PD disaggregated and co-located deployments.

\parahead{Performance model.}
Some prior systems~\cite{agrawal2024taming,goel2026qoserve} use token budgets based on SLO requirements to enable stall-free piggyback execution, modeling batch latency solely as a function of the total number of tokens. However, this approximation is insufficient in agentic workloads, where requests often carry long KV caches. In this regime, attention cost becomes a dominant factor and depends not only on the number of query tokens but also on the KV cache length. Therefore, the scheduler must explicitly model both effects to make reliable admission decisions.

We model the latency of a mixed batch $B$ by decomposing transformer execution into components with distinct scaling behaviors. The key observation is that FFN layers scale with the number of tokens, while attention layers scale with both token count and KV-cache length.

\emph{GEMM.}
All tokens in the batch, including decode tokens and prefill chunk tokens, are processed by the same dense operators in each transformer layer, including the projection matrices in attention (QKV and output projections) and the FFN layers. Let
\[
M = |B_{\mathrm{decode}}| + \sum_{j \in B_{\mathrm{prefill}}} q_j,
\]
denote the total number of forward tokens in the batch. The GEMM cost $T_{\mathrm{GEMM}}(M)$ is non-linear in $M$: for small $M$, execution is memory-bound and exhibits near-constant latency, while for larger $M$, it becomes compute-bound and scales approximately linearly. We therefore model $T_{\mathrm{GEMM}}$ using an offline-profiled lookup table with interpolation.

\emph{Decode attention.}
Each decode request contributes a single query token that attends to its full KV cache of length $L_j$. This operation is memory-bandwidth bound and its cost scales linearly with the total KV tokens accessed:
\[
\alpha_{\mathrm{d}} \sum_{j \in B_{\mathrm{decode}}} L_j,
\]
where $\alpha_{\mathrm{d}}$ is a profiled coefficient.

\emph{Prefill attention.}
Each prefill chunk $j$ contains $q_j$ query tokens attending to a KV cache with prefix length $\mathrm{prefix}_j$. In contrast to decode attention, prefill attention is compute-bound due to the larger number of query tokens per request, and we model its cost based on total attention work. Due to the causal mask, the $i$-th query token attends to $\mathrm{prefix}_j + i + 1$ tokens, leading to
\[
A_j = \sum_{i=0}^{q_j - 1} (\mathrm{prefix}_j + 1 + i)
= q_j \cdot \mathrm{prefix}_j + \frac{q_j(q_j + 1)}{2}.
\]
The total prefill attention cost is therefore
\[
\alpha_{\mathrm{p}} \sum_{j \in B_{\mathrm{prefill}}} A_j,
\]
where $\alpha_{\mathrm{p}}$ is a profiled coefficient.

Combining the three components, the estimated batch latency is
\begin{equation}
\mathrm{EstBatchLatency}(B) =
\begin{array}{l}
\quad T_{\mathrm{GEMM}}(M) \\
\quad + \quad \alpha_{\mathrm{d}} \sum_{j \in B_{\mathrm{decode}}} L_j \\
\quad + \quad \alpha_{\mathrm{p}} \sum_{j \in B_{\mathrm{prefill}}} A_j.
\end{array}
\label{eq:latency-model}
\end{equation}

This model enables accurate estimation of the latency impact of co-batching decode and prefill requests, and is used to guide SLO-aware scheduling decisions.

\parahead{Slack-aware batching for PD disaggregation.}
In PD disaggregated systems~\cite{zhong2024distserve,patel2024splitwise,qin2025mooncake}, prefill and decode are served by separate instances. Lookahead requests correspond to prefill operations and are handled by prefill instances.

\algnotext{EndIf}
\begin{algorithm}[t]
    \caption{Prefill Batch Scheduling (PD Disaggregated)}
    \label{alg:schedule-prefill}
    \begin{algorithmic}[1]
    \renewcommand{\algorithmicrequire}{\textbf{Input:}}
    \renewcommand{\algorithmicensure}{\textbf{Output:}}
    \Require LC queue $\mathcal{Q}_{\mathrm{LC}}$, BE queue $\mathcal{Q}_{\mathrm{BE}}$, LC schedule algorithm $\mathcal{A}_{\mathrm{LC}}$, current time $t_{\mathit{now}}$
    \Ensure batch to be executed $B = B_{\mathrm{LC}} \cup B_{\mathrm{BE}}$

    \vspace{1pt}
    \State \textcolor{blue}{// Schedule LC reqs based on the original algorithm}
    \vspace{1pt}
    \State $\mathcal{Q}_{\mathrm{LC}} \leftarrow \mathrm{Schedule}(\mathcal{Q}_{\mathrm{LC}}, \mathcal{A}_{\mathrm{LC}})$ \label{line:lc-schedule}

    \vspace{6pt}
    \State \textcolor{blue}{// Get minimum slack across LC queue}
    \vspace{1pt}
    \State $p \leftarrow 0$;\quad $s_{\min} \leftarrow +\infty$
    \For{each request $r \in \mathcal{Q}_{\mathrm{LC}}$} \label{line:slack-loop}
        \State $p \leftarrow p + \mathrm{EstPrefillLatency}(r)$
        \State $s_j \leftarrow \bigl(r.t_{\mathit{arrival}} + r.\mathit{SLO}_{\mathit{TTFT}}\bigr) - t_{\mathit{now}} - p$
        \State $s_{\min} \leftarrow \min(s_{\min},\, s_j)$
    \EndFor
    \State $t_{\mathit{budget}} \leftarrow \max(0,\, s_{\min})$ \label{line:budget}

    \vspace{8pt}
    \State \textcolor{blue}{// Construct batch}
    \vspace{2pt}
    \State $B_{\mathrm{LC}} \leftarrow \mathrm{NextChunks}(\mathcal{Q}_{\mathrm{LC}})$ \label{line:init-batch}
    \State $B_{\mathrm{BE}} \leftarrow \emptyset$
    \For{each candidate chunk $c \in \mathcal{Q}_{\mathrm{BE}}$} \label{line:be-loop}
        \If{$\mathrm{EstBatchLatency}(B_{\mathrm{LC}} \cup B_{\mathrm{BE}} \cup \{c\}) > t_{\mathit{budget}}$}
        \State \textbf{break}
        \EndIf
        \State insert $c$ into $B_{\mathrm{BE}}$
    \EndFor

    \vspace{3pt}
    \State \Return $B_{\mathrm{LC}} \cup B_{\mathrm{BE}}$
    \end{algorithmic}
\end{algorithm}

Our goal is to piggyback lookahead requests on prefill instances without violating the TTFT constraints of other requests. To enable fine-grained scheduling, we divide prefill requests into small chunks~\cite{agrawal2024taming,holmes2024deepspeed,qin2025mooncake} and treat each chunk as a schedulable unit. The key idea is to treat lookahead execution as slack-driven computation. Instead of explicitly reserving resources for BE requests, we derive a time budget from the TTFT constraints of LC requests and admit BE chunks only within this budget. This reduces the scheduling problem to a simple question: how much additional latency can be introduced without violating any LC deadline.

As shown in Algorithm~\ref{alg:schedule-prefill}, the scheduler first applies the original LC scheduling policy (line~\ref{line:lc-schedule}), such as first come first serve (FCFS), and then computes the available slack from LC requests. It iterates over the LC queue (line~\ref{line:slack-loop}), accumulating the expected prefill latency of preceding requests, and derives a slack value for each request based on its TTFT deadline. The minimum slack determines the time budget $t_{\mathit{budget}}$ (line~\ref{line:budget}).

Using this budget, the scheduler constructs a batch by first selecting LC chunks (line~\ref{line:init-batch}), and then greedily admitting BE chunks (line~\ref{line:be-loop}). Each candidate chunk is included only if the resulting batch latency does not exceed $t_{\mathit{budget}}$, ensuring that BE execution does not delay any LC request.

\parahead{TBT-constrained batching for PD co-location.}
In PD co-located systems~\cite{agrawal2024taming,holmes2024deepspeed,kamath2025pod}, prefill and decode share the same instance and are executed within a hybrid batch. In this setting, decode requests are subject to TBT SLOs, since each iteration directly affects token generation latency.

Unlike the PD disaggregated setting, where slack is derived from queued requests, TBT constraints in co-located systems are enforced at the granularity of each iteration. Each batch is constructed to satisfy the latency bound $\delta$, and additional work is admitted only if it preserves this constraint. This enables lookahead execution to be interleaved with decode without affecting token generation latency.


To further reduce interference, lookahead prefills are divided into small chunks and interleaved with decode iterations. This chunked execution allows the scheduler to incrementally utilize slack while ensuring that newly arriving LC requests can be promptly scheduled.

\begin{algorithm}[t]
    \caption{Hybrid Batch Scheduling (PD Co-Located)}
    \label{alg:schedule-hybrid}
    \begin{algorithmic}[1]
    \renewcommand{\algorithmicrequire}{\textbf{Input:}}
    \renewcommand{\algorithmicensure}{\textbf{Output:}}
    \Require decode queue $\mathcal{Q}_{\mathrm{dec}}$, LC prefill queue $\mathcal{Q}_{\mathrm{LC}}$, BE prefill queue $\mathcal{Q}_{\mathrm{BE}}$, TBT SLO $\delta$
    \Ensure batch to be executed $B$

    \vspace{1pt}
    \State $B \leftarrow \emptyset$

    \vspace{6pt}
    \State \textcolor{blue}{// Schedule decode requests}
    \vspace{1pt}
    \For{each request $r \in \mathcal{Q}_{\mathrm{dec}}$} \label{line:dec-loop}
        \If{$\mathrm{EstBatchLatency}(B \cup \{r\}) > \delta$} \textbf{break} \EndIf
        \State insert $r$ into $B$
    \EndFor

    \vspace{6pt}
    \State \textcolor{blue}{// Schedule LC prefill requests}
    \vspace{1pt}
    \For{each chunk candidate $c_{LC} \in \mathcal{Q}_{\mathrm{LC}}$} \label{line:lc-loop}
        \If{$\mathrm{EstBatchLatency}(B \cup \{c_{LC}\}) > \delta$} \textbf{break} \EndIf
        \State insert $c_{LC}$ into $B$
    \EndFor

    \vspace{6pt}
    \State \textcolor{blue}{// Schedule BE prefill requests}
    \vspace{1pt}
    \For{each chunk candidate $c_{BE} \in \mathcal{Q}_{\mathrm{BE}}$} \label{line:be-loop-colo}
        \If{$\mathrm{EstBatchLatency}(B \cup \{c_{BE}\}) > \delta$} \textbf{break} \EndIf
        \State insert $c_{BE}$ into $B$
    \EndFor

    \vspace{3pt}
    \State \Return $B$

    \end{algorithmic}
\end{algorithm}

To preserve TBT guarantees, as shown in Algorithm~\ref{alg:schedule-hybrid}, the scheduler constructs the batch in a priority order. It first admits decode requests (line~\ref{line:dec-loop}), ensuring that the core token generation loop is not delayed. It then schedules LC prefill chunks (line~\ref{line:lc-loop}), followed by BE lookahead chunks (line~\ref{line:be-loop-colo}). At each step, a candidate is admitted only if the resulting batch latency remains within the TBT bound $\delta$.

This ordering, combined with the latency constraint, ensures that decode and LC requests are never delayed by BE execution, while allowing the scheduler to exploit slack within each iteration.

\parahead{Discussion on API-based serving.}
The best-effort abstraction for lookahead requests also enables flexible API design. Service providers can expose lookahead requests as a lower-priority class with discounted pricing, since they consume only spare capacity and carry no latency guarantees.
Furthermore, after a lookahead context commits, subsequent generation can directly reuse the cached prefix tokens prepared by lookahead execution, which can be billed at a reduced rate.
This creates a natural incentive for users to adopt lookahead context engineering strategies, aligning system efficiency improvements with reduced serving cost.

\section{Implementation}
\label{sec:implement}

\begin{figure}
    \centering
    \includegraphics[width=0.99\linewidth]{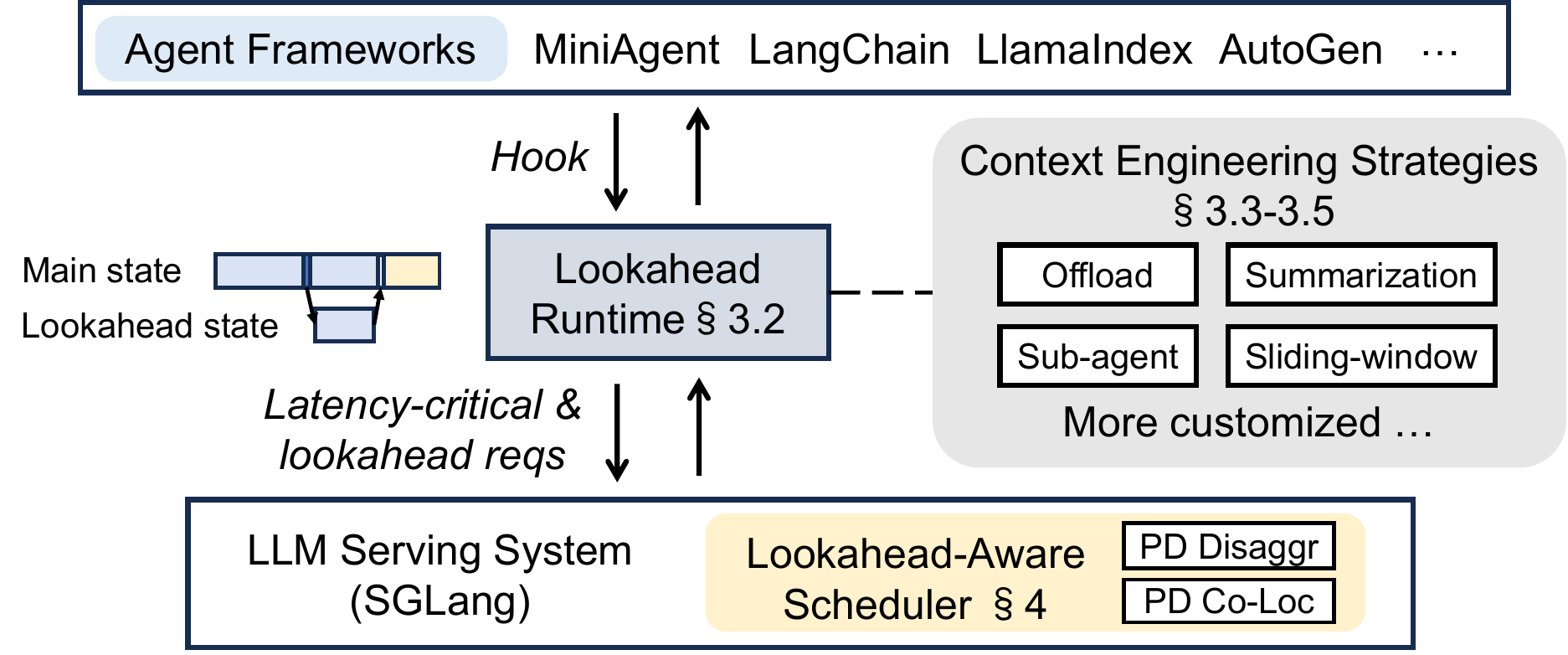}
    \caption{System overview of \SYS{}.}
    \Description{Overview of the BladeAgent runtime between agent frameworks and the LLM serving system.}
    \label{fig:overview}
\end{figure}

We implement \SYS{}, a system that enables lookahead execution for context transformation in LLM-based agents. As illustrated in Figure~\ref{fig:overview}, the \SYS{} runtime sits between agent frameworks (e.g., MiniAgent~\cite{miniagent} and LangChain~\cite{langchain}) and the underlying LLM serving system to orchestrate lookahead execution.

\SYS{} exposes a unified interface for supporting various context engineering strategies, such as offloading, summarization, and sub-agent isolation. Agent frameworks interact with \SYS{} through a simple hook mechanism, allowing lookahead requests to be issued alongside latency-critical requests without modifying application logic. Internally, the runtime of \SYS{} maintains both the main execution state and a separate lookahead state, and manages their interaction through the lookahead programming model introduced in Section~\ref{subsec:programming_model}.

To support lookahead requests, \SYS{} integrates with the LLM serving system, SGLang~\cite{sglang} v0.5.9, and incorporates a lookahead-aware scheduler. The scheduler supports both prefill-decode co-located and disaggregated deployment settings, enabling best-effort execution of lookahead workloads while preserving latency SLO guarantees for foreground requests. This design allows \SYS{} to overlap context transformation with normal inference, effectively hiding overhead from the critical path.

\section{Evaluation}
\label{sec:eval}

\begin{figure*}
    \centering
    \includegraphics[width=0.97\linewidth]{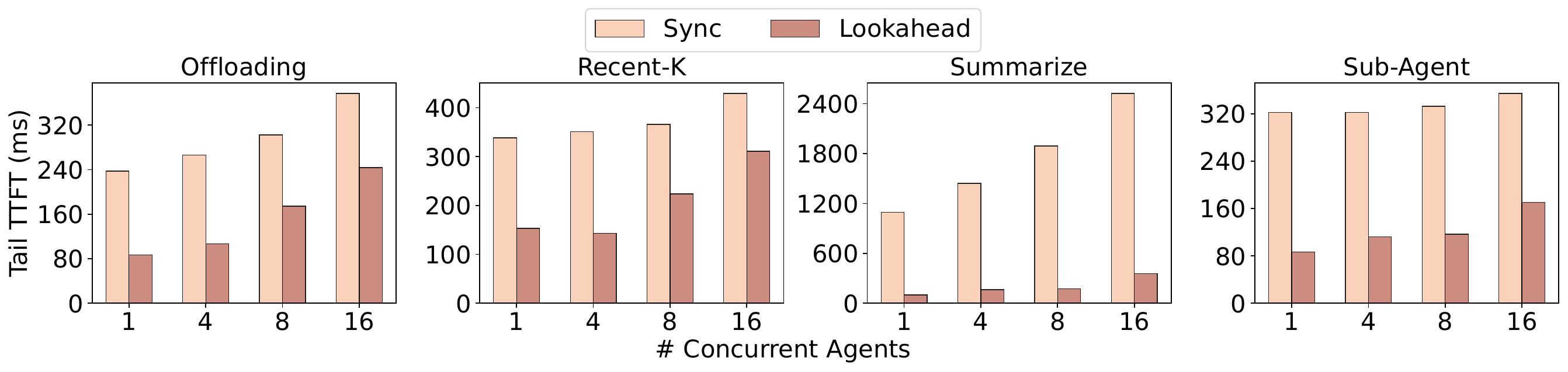}
    \caption{Transform-point TTFT for each strategy on Qwen3-8B (PD co-located) at concurrency levels 1, 4, 8, and 16. \SYS{} eliminates the transformation-induced TTFT spike across all strategies and concurrency levels.}
    \Description{Bar plots of transform-point TTFT for Qwen3-8B under multiple strategies and concurrency levels.}
    \label{fig:e2e_8b}
\end{figure*}

\begin{figure*}
    \centering
    \includegraphics[width=0.97\linewidth]{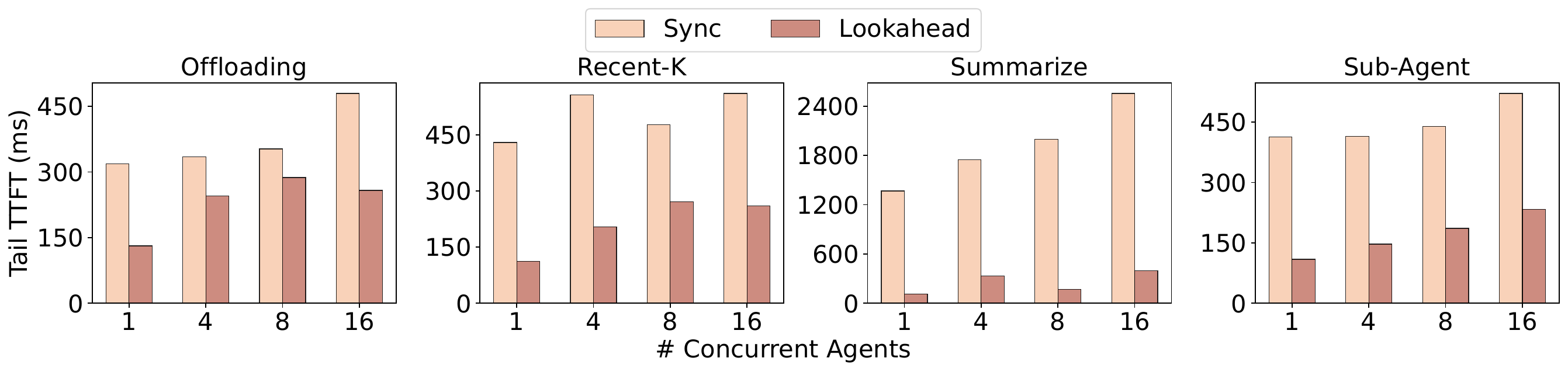}
    \caption{Transform-point TTFT for each strategy on Qwen3-32B (PD co-located, TP=4 across four H100 GPUs).}
    \Description{Bar plots of transform-point TTFT for Qwen3-32B under multiple context engineering strategies.}
    \label{fig:e2e_32b}
\end{figure*}

\subsection{Experimental Setup}

\parahead{Hardware.}
We evaluate \SYS{} on NVIDIA H100 GPUs with 80GB HBM, connected via NVLink within a node. Host machines are equipped with AMD EPYC CPUs.

\parahead{Models.}
We evaluate two representative open-source LLMs: Qwen3-8B and Qwen3-32B~\cite{yang2025qwen3}. Both models use grouped-query attention (GQA)~\cite{touvron2023llama} and dense feed-forward layers, with a maximum context length of 32K tokens.

\parahead{Deployment configurations.}
We evaluate two deployment settings.
In the \textit{PD co-located} setting, prefill and decode share the same GPU instance. We run Qwen3-8B on a single H100 GPU and Qwen3-32B on four H100 GPUs with tensor parallelism (TP=4), varying the number of concurrent agents from 1 to 16. Rather than modeling request arrivals, we adopt a fixed-concurrency configuration: each agent runs a long-lived session with causally interleaved prefill and decode requests, which the arrival-rate-based models for stateless workloads do not capture. Agents are launched at staggered start times to avoid burst arrivals, and each agent runs in an isolated process to eliminate event-loop interference. Due to large KV cache footprints at long contexts, we do not scale concurrency further within a single instance.

In the \textit{PD disaggregated} setting, prefill and decode run on separate instances. Prefill instances serve prefill requests and maintain an SGLang radix cache~\cite{sglang}; decode instances perform token generation. KV caches are transferred between instances via NIXL~\cite{nixl}.

\parahead{Workloads.}
Each agent executes a multi-step code analysis task. At each step, the agent reads source code from the MiniAgent~\cite{miniagent} codebase via shell commands (e.g., \texttt{head}, \texttt{tail}, \texttt{sed}) and produces an analysis response, generating realistic workloads with interleaved LLM calls and tool executions and causing the context to grow monotonically. Each run consists of 28 steps, with approximately 600-650 tokens added per step. All agents receive unique prompts to prevent radix cache sharing from suppressing context growth.

\parahead{Baselines.}
We compare against synchronous context engineering strategies, denoted as \textit{sync}, which are widely adopted in existing agent frameworks. In these systems, context transformations are triggered reactively and executed on the critical path, blocking subsequent LLM generation.
To ensure fair comparison, the baselines implement the same context engineering strategies as \SYS{}, but without lookahead execution. Both the baselines and \SYS{} use SGLang~\cite{sglang} as the LLM serving backend.

\subsection{End-to-End Lookahead Strategy Evaluation}

We implement four context engineering strategies based on MiniAgent~\cite{miniagent}, a lightweight agent framework that enables controlled evaluation without framework-specific optimizations. The strategies are offloading, keep-recent-$K$, summarization, and sub-agent isolation, following the strategy descriptions in Section~\ref{sec:lookahead}. Each strategy is evaluated independently.
We focus on tail TTFT, especially at transformation points where synchronous execution introduces blocking overhead. This allows us to directly evaluate whether lookahead execution removes the latency spikes associated with context transformation.

\parahead{PD co-located.}
Figures~\ref{fig:e2e_8b} and~\ref{fig:e2e_32b} show results on Qwen3-8B and Qwen3-32B in the PD co-located setting.
\SYS{} consistently reduces tail TTFT across all four strategies and both model scales, with an average reduction of 62.0\% and 61.5\% on Qwen3-8B and Qwen3-32B, respectively.

Among these strategies, summarization yields the largest gain, with up to 11.9$\times$ TTFT improvement. Unlike other strategies, summarization incurs both an additional LLM generation to produce the summary and a full re-prefill of the resulting context, leading to substantially higher overhead. \SYS{} moves both operations into the lookahead stream, effectively hiding this compound cost.

In contrast, offloading shows smaller improvements, especially under high concurrency. This is because once earlier segments have been offloaded, subsequent transformations operate on already compact representations and incur little additional transformation cost. As a result, the overhead of synchronous execution diminishes over time, reducing the relative benefit of lookahead.
In practice, offloading alone does not effectively control context growth after repeated triggers, and is often combined with reduction strategies such as summarization~\cite{ji2025manus}.

\begin{figure*}
    \centering
    \includegraphics[width=0.97\linewidth]{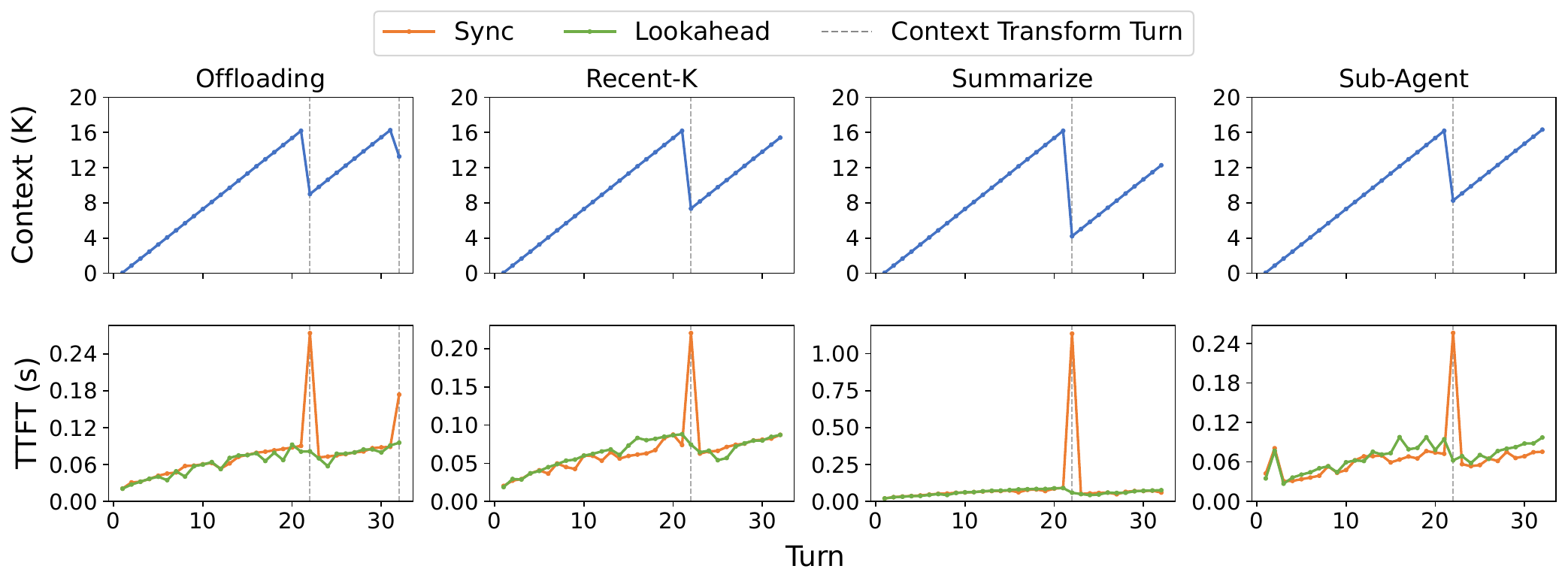}
    \caption{Context length (top) and per-turn TTFT (bottom) for a single Qwen3-8B agent under each strategy. Synchronous transformation produces a TTFT spike at each trigger; \SYS{} absorbs the transformation cost across preceding turns and delivers an uninterrupted TTFT profile.}
    \Description{Time-series plots of context length and per-turn TTFT for a single Qwen3-8B agent.}
    \label{fig:turns}
\end{figure*}

\begin{figure*}
    \centering
\includegraphics[width=0.97\linewidth]{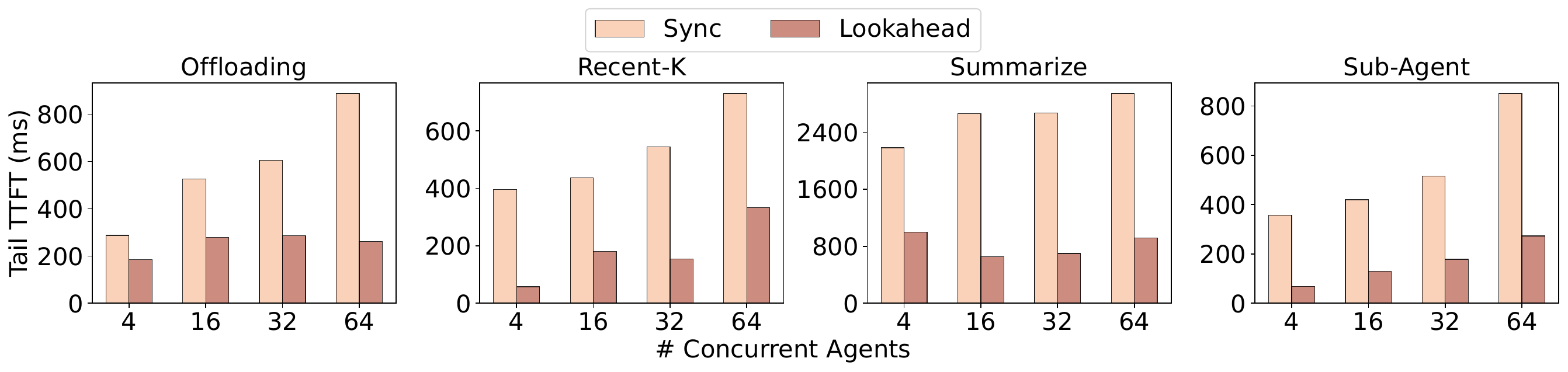}
    \caption{Transform-time TTFT on Qwen3-8B in a PD disaggregated deployment with four prefill and four decode instances.}
    \Description{Transform-time TTFT comparison for Qwen3-8B in a PD disaggregated deployment.}
    \label{fig:e2e_8b_pd}
\end{figure*}

To understand when transformations are triggered and their impact on latency, we plot the context length and TTFT across turns in Figure~\ref{fig:turns} for a single agent on Qwen3-8B.
For offloading, the transformation is triggered when the context exceeds 15K tokens. The first trigger reduces the context size by approximately half, while subsequent triggers yield diminishing reductions as earlier segments have already been offloaded.
For keep-recent-$K$, the transformation is triggered at turn 22, retaining the most recent 10 turns along with the system prompt.
For summarization, the transformation is triggered at 15K tokens, keeping recent turns whose total length exceeds 4K tokens and summarizing the remaining context into 128 tokens. A soft threshold at 11K tokens enables lookahead execution to start summarization ahead of the hard trigger.
For sub-agent, the main agent generates approximately 800 tokens as delegation instructions. The sub-agent includes an additional 800-token system prompt and reads around 6K tokens from the codebase at startup.

As shown in the bottom of Figure~\ref{fig:turns}, all strategies introduce noticeable TTFT spikes at transformation points under synchronous execution, despite reducing context growth. By performing transformations ahead of time and amortizing the cost across multiple turns, \SYS{} eliminates these spikes and produces a smooth TTFT profile.

\parahead{PD disaggregated.}
Figure~\ref{fig:e2e_8b_pd} shows results under PD disaggregation with up to 64 concurrent agents.
We use four H100 GPUs as prefill instances and four H100 GPUs as decode instances to serve Qwen3-8B.
\SYS{} consistently reduces TTFT at transformation points across all strategies, achieving an average reduction of 64.5\%. Compared to the co-located setting, both the baseline and \SYS{} exhibit higher absolute TTFT due to additional KV cache transfer and routing overhead.
Despite this overhead, lookahead execution remains effective and continues to eliminate transformation-induced latency spikes.

\parahead{Agent framework integration.}
We further integrate \SYS{} into representative agent frameworks, including LangChain~\cite{langchain} (turn/token-based sliding window and summarization), LlamaIndex~\cite{Liu_LlamaIndex_2022} (sliding window and summarization), AutoGen~\cite{autogen} (sliding window), and OpenClaw~\cite{openclaw} (summarization). These strategies correspond to the context engineering mechanisms natively supported in each framework. In particular, sliding window is another form of context truncation. It maintains a fixed-size (turns or tokens) context by continuously evicting old tokens at each turn, whereas keep-recent-$K$ performs truncation only when the context exceeds a predefined limit, resulting in periodic rather than continuous transformations.
We replace these built-in strategies with their lookahead-optimized counterparts without changing the agent logic, and evaluate them under the same workloads. Across these frameworks, \SYS{} consistently reduces transformation-point TTFT by 26.8\% to 80.5\%.

\subsection{Lookahead-Aware Scheduler Evaluation}

\begin{figure}
    \centering
    \includegraphics[width=0.97\linewidth]{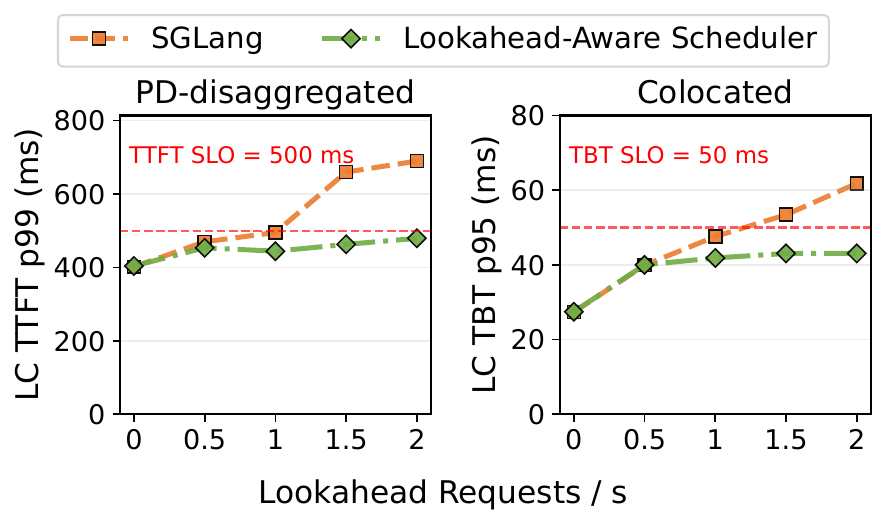}
    \caption{Impact of lookahead traffic on latency-critical requests. Our lookahead-aware scheduler effectively reduces the interference as lookahead traffic increases.}
    \Description{Latency-critical request latency under increasing best-effort lookahead traffic.}
    \label{ig:be_lc_interference}
\end{figure}

We evaluate whether the lookahead-aware scheduler mitigates interference from best-effort (BE) lookahead traffic on latency-critical (LC) requests.
We keep the LC workload fixed, vary the intensity of BE lookahead traffic (i.e., number of lookahead requests per second), and compare \SYS{} with vanilla SGLang, whose scheduler admits BE requests without lookahead-aware control.
As shown in Figure~\ref{ig:be_lc_interference}, the uncontrolled policy causes LC latency to increase rapidly as lookahead load grows.
In the PD-disaggregated setting, LC TTFT p99 exceeds the target under high BE traffic; in the co-located setting, LC TBT rises sharply because BE chunks share the same execution loop as decoding.
\SYS{} avoids this behavior by prioritizing LC requests and admitting BE lookahead chunks only when the predicted batch latency fits within the available slack.
This keeps LC latency substantially lower across the same BE load range, showing that lookahead-aware admission control allows BE work to make progress without dominating the latency-critical path.

\begin{figure}
    \centering
    \includegraphics[width=0.97\linewidth]{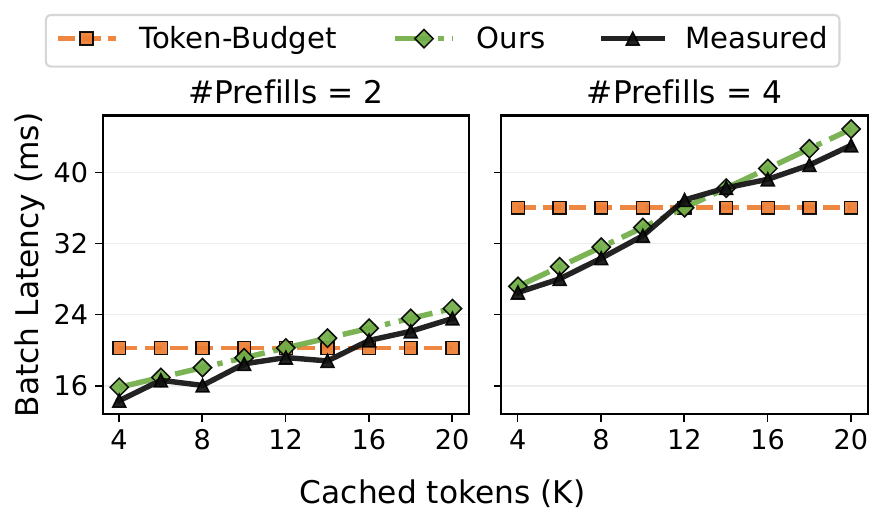}
    \caption{Accuracy of the context-aware performance model from Section~\ref{sec:colocate} under controlled mixed batches with fixed generation work and varying cached-context lengths.}
    \Description{Line plots comparing measured batch latency with token-budget-only and context-aware predictions as cached-context length increases.}
    \label{fig:prediction_error}
\end{figure}

We also evaluate the context-aware performance model introduced in Section~\ref{sec:colocate}, which predicts mixed-batch latency from both the number of forward tokens and the KV-cache lengths accessed by decode and prefill attention. 
We manually construct batches with four decoding requests and either two or four prefill chunks, where each prefill chunk contains 256 tokens. The amount of new generation work in each batch is fixed, while we vary the number of cached tokens already accumulated by each request. As shown in Figure~\ref{fig:prediction_error}, across the plotted two- and four-prefill-chunk configurations, the token-budget-only model reaches a maximum relative error of 41.5\%, while our performance model keeps the maximum relative error to 13.7\%.
This gap shows that a fixed token budget alone is insufficient: batches with the same amount of forward-token work can have different execution times when their cached contexts differ.
By explicitly modeling KV-cache length in both decode and prefill attention, \SYS{} captures the hidden attention-cost growth caused by long agent contexts and gives the scheduler a more reliable signal for admitting lookahead chunks without violating LC latency constraints.

\section{Related Work}
\label{sec:related}

\parahead{LLM serving systems.}
A large body of work has focused on optimizing general-purpose LLM serving systems.
vLLM~\cite{vllm} and SGLang~\cite{sglang} are among the most widely used open-source serving engines, and they incorporate a range of important optimizations, including continuous batching~\cite{yu2022orca}, paged attention~\cite{vllm}, and prefix sharing~\cite{sglang}.
Chunked prefill techniques~\cite{agrawal2024taming,holmes2024deepspeed} split prefilling into smaller chunks and mix them with decoding, preventing long prefills from blocking decoding TBT while improving GPU utilization in prefill-decode co-located settings.
Prefill-decode disaggregation~\cite{zhong2024distserve,patel2024splitwise} eliminates interference between prefill and decode requests by dispatching them to separate instances, and has become the de facto design in large-scale datacenter LLM serving.
More recent work~\cite{zhu2025megascale,wang2025step} further proposes disaggregating attention and FFN computation in MoE models due to their mismatched computational characteristics, enabling decoupled scaling and more effective use of heterogeneous hardware.
Our work optimizes the request scheduler in serving engines to support our lookahead programming model under both PD co-located and disaggregated deployments.

Lots of work~\cite{qin2025mooncake,liu2025lmcache,yao2025cacheblend,gao2025apt,lee2024infinigen,pan2025database} study efficient KV cache management to improve system performance.
Prefix caching~\cite{sglang,gim2024prompt,gao2024cost,yu2025stateful} stores KV caches on GPUs or external storage after requests complete, avoiding recomputation when subsequent requests share the same prefix. This capability is particularly important in workloads such as multi-turn conversations. Our lookahead execution also relies on prefix caching to preserve the KV cache of transformed contexts.
Another line of work~\cite{dao2022flashattention,ye2025flashinfer,dong2024flex,pan2025fasttree,kamath2025pod,sanovar2025leanattention,wu2025mirage,cheng2025mirage,zheng2023bladedisc,xia2023flash} focuses on GPU kernel optimizations for LLM serving.
A representative example is FlashInfer~\cite{ye2025flashinfer}, which provides efficient kernels for a variety of serving scenarios and has been integrated into many serving frameworks.
These kernel-level optimizations are orthogonal to our work.

\parahead{Agent serving optimization.}
As agentic workloads become increasingly important, recent research has begun to explore system-level optimizations for agent serving.
Parrot~\cite{lin2024parrot} and Ayo~\cite{tan2025towards} model static agent workflows as directed acyclic graphs (DAGs) and apply a range of optimizations over the execution graph.
In contrast, our work does not assume a static workflow and instead targets more autonomous, multi-turn agents.
Other lines of work focus on retrieval-augmented generation (RAG)~\cite{hu2025hedrarag,lin2025telerag,jin2025ragcache,yao2025cacheblend,hu2026pancake,xu2025tribase,xu2025harmony,yu2025aquapipe,jiang2024chameleon} and on multi-agent simulation~\cite{xie2025ai,pan2026scalesim}, both of which differ from our target setting.
Autellix~\cite{luo2025autellix} mitigates head-of-line blocking by scheduling at the granularity of agent programs rather than individual requests.
InferCept~\cite{abhyankar2024infercept}, KVFlow~\cite{pan2025kvflow}, and Helium~\cite{wadlom2026efficient} improve KV cache management for agentic workloads by adopting application-aware eviction and fetching policies instead of conventional LRU.
ThunderAgent~\cite{kang2026thunderagent} and CONCUR~\cite{chen2026concur} identify memory thrashing in multi-turn agents and propose admission control mechanisms based on current system load.
While these works optimize agent workloads from different perspectives, they do not address inefficiencies arising from context engineering, which is the focus of our work.

\section{Conclusion}

In this paper, we identify context transformation overhead as a key bottleneck in LLM-based agent systems, arising from KV cache invalidation and recomputation under common context engineering strategies.
By observing that these transformations are segment-decomposable, we enable their execution ahead of time. We propose a lookahead programming model that expresses transformations as asynchronous operations, together with a runtime and scheduling support to execute them proactively without impacting latency-critical requests.
Our results show that this approach effectively eliminates transformation overhead and significantly improves TTFT.

\bibliographystyle{ACM-Reference-Format}
\bibliography{ref}

\end{document}